# A Comparative Study Between a Classical and Optimal Controller for a Quadrotor

Prathamesh Saraf
Department of Electrical &
Electronics Engineering
Birla Institute of Technology and
Science – Pilani
Hyderabad, India
pratha1999@gmail.com

Manan Gupta
Department of Electrical &
Electronics Engineering
Birla Institute of Technology and
Science – Pilani
Hyderabad, India
f20170488@hyderabad.bits-pilani.ac.in

Dr. Alivelu Manga Parimi
Department of Electrical &
Electronics Engineering
Birla Institute of Technology and
Science – Pilani
Hyderabad, India
alivelu@hyderabad.bits-pilani.ac.in

*Abstract*— This paper presents a simulation-based comparison between the two controllers, Proportional Integral Derivative (PID), a classical controller and Linear Quadratic Regulator (LQR), an optimal controller, for a linearized quadrotor model. To simplify an otherwise complicated dynamic model of a quadrotor, we derive a linear mathematical model using Newtonian and Euler's laws and applying basic principles of physics. This derivation gives the equations that govern the motion of a quadrotor, both concerning the body frame and the inertial frame. A state-space model is developed, which is then used to simulate the control algorithms for the quadrotor. Apart from the classic PID control algorithm, LQR is an optimal control regulator, and it is more robust for a quadrotor. Both the controllers are simulated in Simulink under the same initial conditions and show a satisfactory response.

## I. INTRODUCTION

Quadcopters (unmanned aerial vehicles powered by four rotors) are finding increasing applications in fields ranging from video surveillance to emergency response. They are required to fly in unknown environments and must do so with accurate control. Controllers for quadcopters have been studied for quite some time now. Most controllers used on quadcopters utilize classical control, such as PID control. A proportional-integral-derivative (PID) controller is a control mechanism very commonly used in most quadcopter systems. Classical control theory encompasses linear time-invariant single-input single-output systems. The basis of control for these systems depends on how their behavior is modified using a feedback loop. On the other hand, another commonly used controller is the LQR, a type of optimal control. Optimal control theory focuses on mathematical optimization of an objective cost function for a dynamic system.

Shahida et al. [1] have explored how PID controllers give better stability by bringing the closed-loop poles to the more negative side of the s-plane as compared to LQR controllers. Hayrettin et al. [2] tackle quadcopter control's problem by proposing an adaptive controller that is a hybrid of both PID and LQR controllers. Meera et al. [3] present results on a PID controller coupled with a Kalman filter being applied to UAVs. Lucas et al. [4] work on using LQR control to tune a PID controller, combining the control techniques to control the quadcopter.

Existing research in the field has stated how PID has successfully been implemented for control in simulations and even real-world scenarios. However, when we extend the use cases to scenarios where external factors can cause considerable disturbances to a quadcopter's flight, we are motivated to study optimal control to see how it can factor in such uncertain circumstances.

LQR is an optimal control regulator and is expected to be more robust for a quadcopter. LQR focuses on non-linear models rather than the classical linear equation approach of PID. The main drawback of PID controllers is that every test on the actual system requires its linearization. For LQR control, this step is not needed, and one can directly feed in the system equations to the controller and get the desired response.

This study aims to simulate both classical and optimal controllers and obtain results for a comparison of their performances. The feasibility of implementing an LQR controller for a quadcopter is studied. We also believe this work can soon be extended to other unmanned aerial vehicles such as fixed-wing aircraft and vertical take-off and landing (VTOL) vehicles, which open up a wide avenue of applications.

The following sections present the modeling and simulation, followed by the results. In section II on the Quadcopter Model, we first detail the quadcopter's physical model and layout the equations governing its motion. These equations are linearized, and a state-space model is developed in the third section. The fourth and fifth sections follow up with a comprehensive study and observations of the PID and LQR controllers. The Results section contains various cases for flight simulation using the two controllers. These are compared, and the conclusion is presented.

## II. QUADROTOR MODEL

The position of a quadrotor in space can be specified using two reference frames: fixed and mobile. The fixed (inertial) coordinate system is one where Newton's laws are valid and it is with respect to the Earth. The second frame is with respect to the center of the quadrotor frame, where the flight controller is positioned. The Quadrotor has four Degrees of Freedom, namely Thrust, Roll, Pitch, and Yaw. The position and movement of the quadrotor are controlled by varying the speeds and rotation direction of the four motors (two diametrically opposite motors spin clockwise and the other two motors spin anti-clockwise). However, the quadrotor has four inputs (4 motors) and 6 outputs (three translational and three rotational motions along the three axes), and hence, it is an underactuated nonlinear complex system.

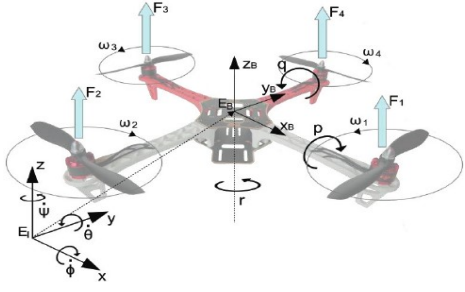

Fig. 1: Quadrotor illustration with inertial frames and 6 DOF coordinates. $\omega_i$ is the angular speed and $F_i$ is the propeller force of the i[th] rotor.

The Euler angles $\varphi \in [-\pi, \pi]$, $\theta \in [\pi/2, \pi/2]$, $\psi \in [-\pi, \pi]$ are used to give the orientation of the mobile reference frame with respect to the inertial frame. These angles help form a rotation matrix that is used to represent rotation/orientation in the same way position is used to represent displacement vectors. The non-linear equations of the quadrotor system are given as [5]:

$$\ddot{x} = (cos\varphi sin\theta cos\psi + sin\varphi sin\psi)\frac{F}{m} \quad (1)$$

$$\ddot{y} = (cos\varphi sin\theta sin\psi - sin\varphi cos\psi)\frac{F}{m} \quad (2)$$

$$\ddot{z} = -g + (cos\varphi cos\theta)\frac{F}{m} \quad (3)$$

$$\dot{p} = \frac{I_{yy} - I_{zz}}{I_{xx}}qr - \frac{Jr}{I_{xx}}q\omega + \frac{u_2}{I_{xx}} \quad (4)$$

$$\dot{q} = \frac{I_{zz} - I_{xx}}{I_{yy}}pr - \frac{Jr}{I_{yy}}p\omega + \frac{u_3}{I_{yy}} \quad (5)$$

$$\dot{r} = \frac{I_{xx} - I_{yy}}{I_{zz}}pq + \frac{u_4}{I_{zz}} \quad (6)$$

where $Jr$ is the rotor inertia and p, q, and, r represents the angular velocities with respect to roll, pitch and yaw motions respectively. The input forces are given as:

$$u_1 = K_f * (\omega_1^2 + \omega_2^2 + \omega_3^2 + \omega_4^2) \quad (7)$$

$$u_2 = K_f * (\omega_4^2 - \omega_2^2) \quad (8)$$

$$u_3 = K_f * (\omega_1^2 - \omega_3^2) \quad (9)$$

$$u_4 = K_m * (\omega_1^2 - \omega_2^2 + \omega_3^2 - \omega_4^2) \quad (10)$$

where $K_f$ is the thrust factor and $K_m$ is the drag factor whose value depends on propeller size and air conditions in which the quadrotor is flying, $\omega_i$ is the angular speed of the i[th] rotor. The quadrotor parameters considered for simulation are given in table 1.

TABLE I.

| Sr. No. | Parameter | Value |
|---|---|---|
| 1. | Quadrotor mass ($m$) | 1 Kg |
| 2. | Arm length ($l$) | 22.5cm |
| 3. | Thrust Factor ($K_f$) | 9.8×10[-6] |
| 4. | Drag Factor ($K_m$) | 1.6×10[-7] |
| 5. | $I_{xx}$ | 0.0035 |
| 6. | $I_{yy}$ | 0.0035 |
| 7. | $I_{zz}$ | 0.005 |

The LQR optimization problem requires a linear state-space system as the A and B matrices along with the Q and R optimal control matrices are necessary for computing the full state feedback K matrix. The following section focusses on modelling the non-linear quadrotor equations as a linear state-space system.

### III. STATE-SPACE MODELING

The quadrotor state-space system consists of 12 state variables, four input variables, and four output variables. The state variables represent the absolute quadrotor orientation in space consisting of linear and rotational coordinates and their respective velocities. The input variables consist of the four quadrotor motions, namely thrust, roll, pitch, and yaw motion. The output consists of the required state variables for stability analysis of the quadrotor, which are the vertical displacement (z), roll ($\varphi$), pitch ($\theta$), and yaw ($\psi$) angle displacements.

$$\dot{X} = AX + BU \quad (11)$$

$$Y = CX + DU \quad (12)$$

$$X^T = [x \quad y \quad z \quad \varphi \quad \theta \quad \psi \quad \dot{x} \quad \dot{y} \quad \dot{z} \quad p \quad q \quad r] \quad (13)$$

$$U^T = [u_1 \quad u_2 \quad u_3 \quad u_4] \quad (14)$$

$$Y^T = [z \quad \varphi \quad \theta \quad \psi] \quad (15)$$

where A is a 12×12 state matrix and B is a 12×4 input matrix which are given by [5]:

$$A = \begin{bmatrix} 0 & 0 & 0 & 1 & 0 & 0 & 0 & 0 & 0 & 0 & 0 & 0 \\ 0 & 0 & 0 & 0 & 1 & 0 & 0 & 0 & 0 & 0 & 0 & 0 \\ 0 & 0 & 0 & 0 & 0 & 1 & 0 & 0 & 0 & 0 & 0 & 0 \\ 0 & 0 & 0 & 0 & 0 & 0 & \frac{u_1}{m} & 0 & 0 & 0 & 0 & 0 \\ 0 & 0 & 0 & 0 & 0 & 0 & -\frac{u_1}{m} & 0 & 0 & 0 & 0 & 0 \\ 0 & 0 & 0 & 0 & 0 & 0 & 0 & 0 & 0 & 0 & 0 & 0 \\ 0 & 0 & 0 & 0 & 0 & 0 & 0 & 0 & 0 & 1 & 0 & 0 \\ 0 & 0 & 0 & 0 & 0 & 0 & 0 & 0 & 0 & 0 & 1 & 0 \\ 0 & 0 & 0 & 0 & 0 & 0 & 0 & 0 & 0 & 0 & 0 & 1 \\ 0 & 0 & 0 & 0 & 0 & 0 & 0 & 0 & 0 & 0 & 0 & 0 \\ 0 & 0 & 0 & 0 & 0 & 0 & 0 & 0 & 0 & 0 & 0 & 0 \\ 0 & 0 & 0 & 0 & 0 & 0 & 0 & 0 & 0 & 0 & 0 & 0 \end{bmatrix} \quad (16)$$

$$B = \begin{bmatrix} 0 & 0 & 0 & 0 \\ 0 & 0 & 0 & 0 \\ 0 & 0 & 0 & 0 \\ 0 & 0 & 0 & 0 \\ 0 & 0 & 0 & 0 \\ 0 & 0 & 0 & 0 \\ \frac{1}{m} & 0 & 0 & 0 \\ 0 & 0 & 0 & 0 \\ 0 & 0 & 0 & 0 \\ 0 & \frac{1}{I_{xx}} & 0 & 0 \\ 0 & 0 & \frac{1}{I_{yy}} & 0 \\ 0 & 0 & 0 & \frac{1}{I_{zz}} \end{bmatrix} \quad (17)$$

C matrix is a 4×12 matrix with each column having a unity multiplier to scrape out the 4 state parameters from X, as mentioned above. D is a 4×4 null matrix. Once the state-space equations are obtained, the controller is designed for the quadrotor system. The following section describes the action of a PID controller on the quadrotor system.

## IV. PROPORTIONAL INTEGRAL DERIVATIVE CONTROL

### A. Overview

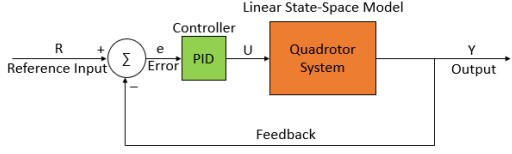

Fig.2: The closed loop quadrotor system with PID controller for stabilization

The PID controller is one of the standard classic models in control theory. The aim is to minimize the error by tuning the proportional, integral, and the derivative coefficients used in the controller equation. Error is defined as the difference between the setpoint value and the actual controller output at any particular instance. The proportional term increases/decreases the error by multiplying with a proportionality constant. The derivative term is used to estimate the controller's future response based on the error with respect to time. The derivative constant also accounts for the damping factor and needs to be tuned to minimize damping and obtain a smooth response. The integral term in the controller sums up all the past values, and the integral coefficient is used for assigning the weight to the integral term to obtain the desired output response in the shortest possible time. Thus, all three coefficients need to be tuned to obtain the setpoint value of the control system.

### B. Equations

The PID control equation uses the $K_p$, $K_i$, and, $K_d$ parameters and the error value to compute the controller response. The control equation is given below:

$$U(t) = K_p e(t) + K_i \int_0^t e(t)dt + K_d \frac{de(t)}{dt} \quad (18)$$

$K_d$ and $K_i$ are defined as:

$$K_d = K_p T_d \ \& \ K_i = K_p/T_i \quad (19)$$

where $T_d$ and $T_i$ are the time periods for integral and derivative actions. The error value is defined as:

$$e(t) = setpoint\ value - actual\ value \quad (20)$$

### C. Controller Design

In the Quadrotor system, the state parameters need to be controlled so that it remains stable during the flight period. A small deviation in the setpoint values of the quadrotor would create large variations in motion, which would lead to deviation from its desired trajectory and making it difficult for the user to control. Even if the main control lies in the hand of the user flying the quadrotor, unavoidable disturbances during the time of flight will create instability and malfunction of the system. Thus, there is a need for an internal flight controller, which would autonomously adjust the rotor to a stable position. The four inputs thrust, roll, pitch, and yaw motion control the six-state parameters depending on the action. Thrust controls the $z$ parameter, roll controls the $x$ and the $\theta$ parameters, pitch controls $y$ and $\varphi$ parameter and yaw controls the $\psi$ angle [6]. Thus, the six-state parameters require 6 PID controllers with 1-2-2-1 controllers for thrust, roll, pitch, and yaw motion control, respectively. For roll and pitch motion, a cascaded PID loop is designed to have an inner and an outer feedback loop. The inner loop needs to be faster than the outer loop for an efficient response of the controller. Slowing the outer loop increases the overshoot but reduces the setting time for the rotational coordinates and their velocity and increases the time for the corresponding translational coordinate to attain the set point. The tuned PID controller parameters for optimal control are given in table 2 below.

TABLE II.

| Sr. No | Input Action | Parameter | Inner Loop | Outer Loop |
|---|---|---|---|---|
| 1. | Thrust | P | 9.09 | NA. |
|  |  | I | 1.94 |  |
|  |  | D | 10.41 |  |
| 2. | Roll | P | 4.04 | -2.92 |
|  |  | I | 10.03 | -0.032 |
|  |  | D | 0.33 | -4.68 |
| 3. | Pitch | P | 4.04 | -2.92 |
|  |  | I | 10.03 | -0.032 |
|  |  | D | 0.33 | -4.68 |
| 4. | Yaw | P | $1.3 \times 10^{-2}$ | NA. |
|  |  | I | $7.6 \times 10^{-4}$ |  |
|  |  | D | $4.9 \times 10^{-2}$ |  |

After the successful implementation of the PID controller, the optimal LQR controller is designed for the linearized quadrotor model.

## V. LINEAR QUADRATIC REGULATOR

### A. Overview

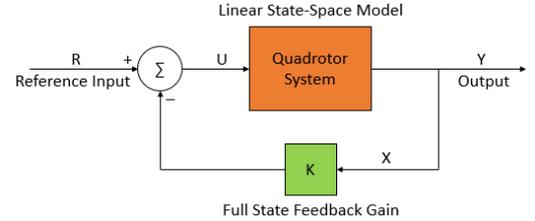

Fig.3: The closed loop quadrotor system with LQR controller for stabilization

As compared to the classic PID controller, the LQR controller involves a lot of mathematical computations to calculate the full state feedback matrix K. LQR controller uses an optimal control algorithm to minimize the cost function defined by the system equations. The cost function involves the state parameters and the input parameters to the system along with the Q and R matrices. The overall cost function needs to be minimum for optimal LQR solution. The Q and R matrices represent the weights assigned to the state parameters and the input parameters. By varying the values of the two matrices, the total value of the cost function can be adjusted according to the desired output. The two main quantities that need to be optimized for the quadrotor model are the power consumption and response speed. For a faster response of the controller, the Q matrix values need to be changed, whereas for minimizing the power consumption while achieving the desired setpoint without focusing on the time of response, the R matrix values need to be adjusted [8].

## B. Equations

The LQR optimization problem requires a linearized state-space model of the system. The cost function which needs to be optimized is given by:

$$J = \int (X^T Q X + U^T R U) dt \quad (21)$$

After tuning the matrices according to the desired output matrix, the Q and R matrices are used to solve the Algebraic Riccati Equation (ARE) to compute the full state feedback matrix, which is essentially the LQR controller.

$$A^T S + SA - SBR^{-1}B^T S + Q = 0 \quad (22)$$

The S matrix obtained from the ARE is used to calculate the Full state feedback gain matrix K using the equation:

$$K = R^{-1} B^T S \quad (23)$$

The final feedback control relation is then calculated as:

$$U = -K * X \quad (24)$$

The tuned Q and R matrices which give the optimal response for the Quadrotor system parameters as defined before are:

$$Q = \begin{bmatrix} 500 & 0 & 0 & 0 & 0 & 0 & 0 & 0 & 0 & 0 & 0 & 0 \\ 0 & 200 & 0 & 0 & 0 & 0 & 0 & 0 & 0 & 0 & 0 & 0 \\ 0 & 0 & 1.71 & 0 & 0 & 0 & 0 & 0 & 0 & 0 & 0 & 0 \\ 0 & 0 & 0 & 500 & 0 & 0 & 0 & 0 & 0 & 0 & 0 & 0 \\ 0 & 0 & 0 & 0 & 200 & 0 & 0 & 0 & 0 & 0 & 0 & 0 \\ 0 & 0 & 0 & 0 & 0 & 3 & 0 & 0 & 0 & 0 & 0 & 0 \\ 0 & 0 & 0 & 0 & 0 & 0 & 10 & 0 & 0 & 0 & 0 & 0 \\ 0 & 0 & 0 & 0 & 0 & 0 & 0 & 10 & 0 & 0 & 0 & 0 \\ 0 & 0 & 0 & 0 & 0 & 0 & 0 & 0 & 10 & 0 & 0 & 0 \\ 0 & 0 & 0 & 0 & 0 & 0 & 0 & 0 & 0 & 0.25 & 0 & 0 \\ 0 & 0 & 0 & 0 & 0 & 0 & 0 & 0 & 0 & 0 & 10 & 0 \\ 0 & 0 & 0 & 0 & 0 & 0 & 0 & 0 & 0 & 0 & 0 & 1 \end{bmatrix} \quad (25)$$

$$R = \begin{bmatrix} 1 & 0 & 0 & 0 \\ 0 & 0.001 & 0 & 0 \\ 0 & 0 & 0.001 & 0 \\ 0 & 0 & 0 & 0.001 \end{bmatrix} \quad (26)$$

And the Feedback gain matrix K is computed as:

$$K = 10 * \begin{bmatrix} 0 & 0 & 0.13 & 0 & 0 & 0.29 & 0 & 0 & 0 & 0 & 0 & 0 \\ 0 & -44.7 & 0 & 0 & -84.8 & 0 & 12.6 & 0 & 0 & 1.6 & 0 & 0 \\ 70.7 & 0 & 0 & 151.6 & 0 & 0 & 0 & 27.5 & 0 & 0 & 10 & 0 \\ 0 & 0 & 0 & 0 & 0 & 0 & 0 & 0 & 10 & 0 & 0 & 3.2 \end{bmatrix} \quad (27)$$

The next section presents the detailed comparative analysis of the two controllers tested for different input conditions.

## VI. RESULTS AND DISCUSSION

The PID and the LQR controller is designed for the quadrotor system in MATLAB as shown in Fig.2 and Fig.3. The simulation tests are divided into 3 cases that help verify the controller functionality in detail. Each case either uses different initial conditions or defines another set-point for a particular parameter of the system. The cases are defined as:

1. Pure thrust input with zero initial conditions
2. Pure thrust input with non-zero initial conditions
3. Thrust and Yaw input with zero initial conditions

Initial conditions define the Quadrotor state just before the controller action, and the set-point defines the desired final steady-state of the system. The detailed analysis of each case along with controlller response is given below:

### A. Case 1

The initial conditions of the Quadrotor are set as zero for all the state parameters and pure thrust action is fed as the system input. In this case, only the vertical displacement ($z$) is expected and should reach a steady state position once the thrust force balances the gravitational force. The vertical velocity ($\dot{z}$) increases rapidly in the beginning and slows down to zero as the rotor achieves steady position midair.

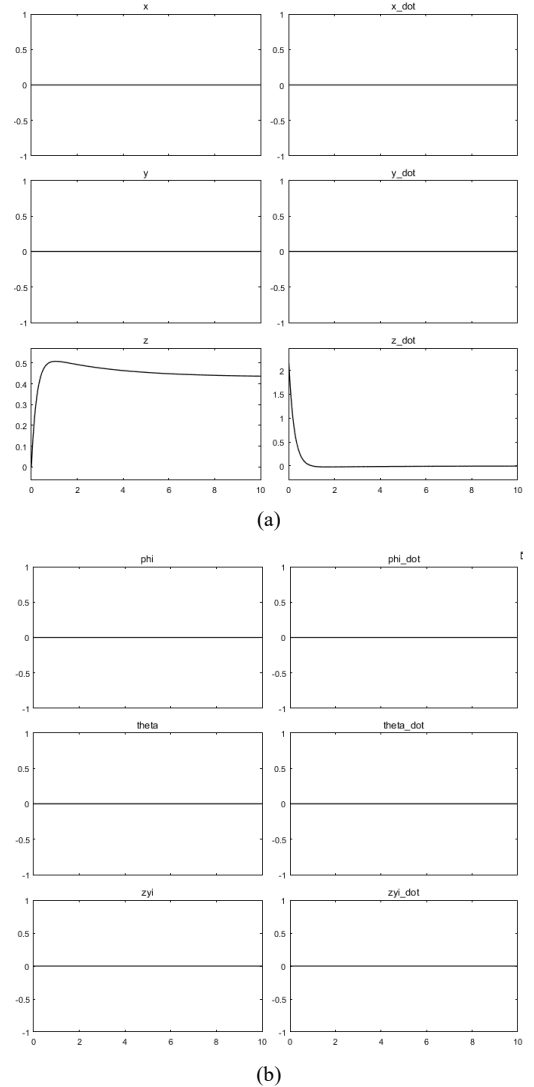

Fig.4: The translational and rotational response of the quadrotor for a PID controller for case 1.

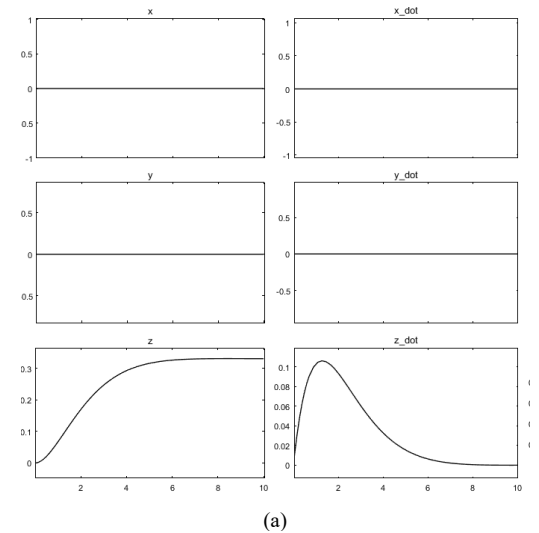

(a)

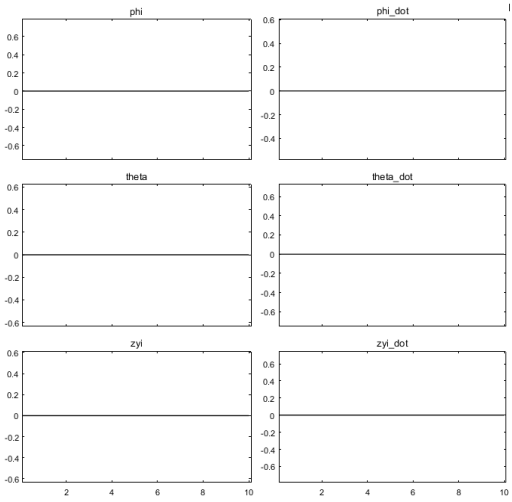

(b)

Fig.5: The translational and rotational response of the quadrotor for an LQR controller for case 1.

Fig.4 and Fig.5 shows the simulated response for PID and LQR controllers respectively. It can be seen that z response for PID controller gives an overshoot of 0.5units and settles at 0.45units with a settling time of 0.8secs whereas there is no overshoot in case of LQR controller and the value settles at 0.7secs at 0.34 units. No fluctuations about the setpoint describes shows that LQR provides a stable output compared to PID controller.

B. *Case 2*

The initial conditions for the system are set as:

$$[1 \quad 1 \quad 0.2 \quad 1 \quad 1 \quad 0 \quad 1 \quad 1 \quad 1 \quad 1 \quad 1 \quad 1]$$

Where each value defines the position of each of the state parameter. The input is again given as a pure thrust force. For this case, all the state parameters except z should achieve the set point which is zero as no other input rotor action is desired. The initial conditions mentioned, show that the rotor is deviated from its set-point and thus the controller will act upon to bring the rotor back to its set-point-zero. The PID and LQR controller responses are shown in Fig. 6 and Fig. 7

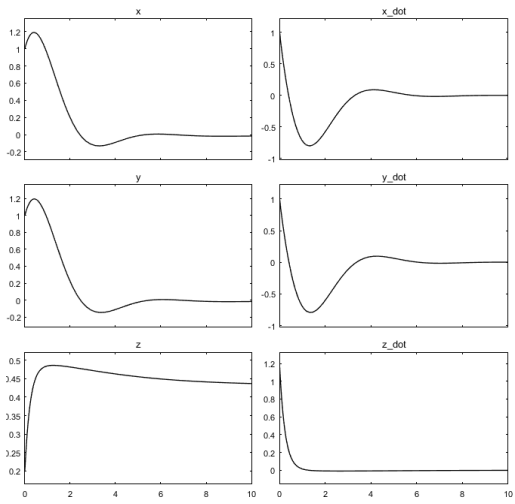

(a)

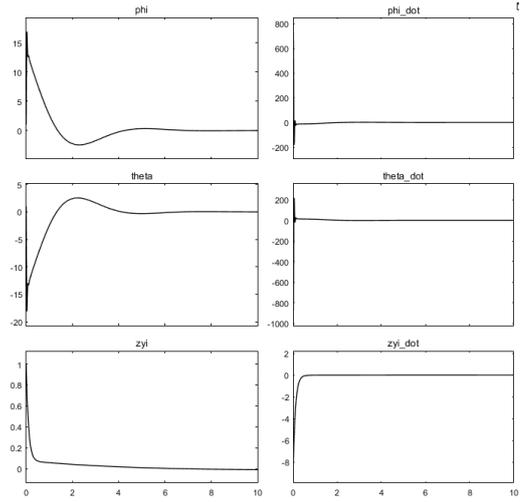

(b)

Fig.6: The rotational response of the quadrotor for a PID controller for case 2.

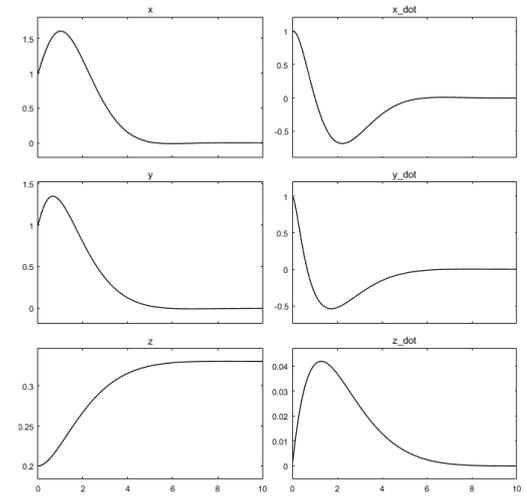

(a)

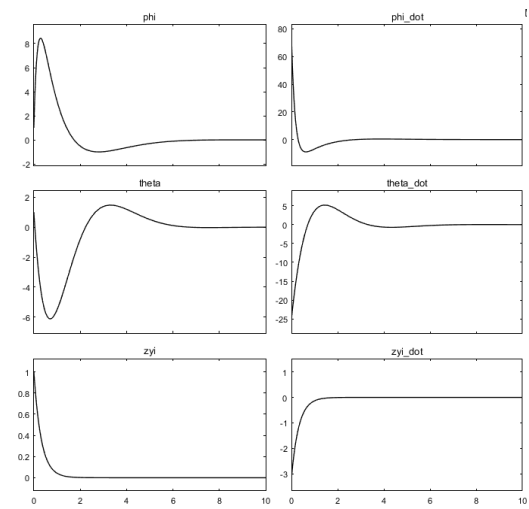

(b)

Fig.7: The translational and rotational response of the quadrotor for an LQR controller for case 2.

Both the controllers show similar settling time of 7 secs for all the state parameters. Although, the 1st overshoot of PID controller (1.2 units) is lesser than LQR controller (1.6 units) for x and y parameters, the PID controllers gives 2nd overshoot in its response in most of the state parameters as opposed to a single overshoot peak in case of an LQR controller. More the number of overshoot peaks, lesser is the stability. Thus, LQR response is more stable and robust as compared to the PID controller.

*C. Case 3*

The initial conditions are set to zero and a thrust force along with a yaw motion is given as an input to both the controllers. A desired yaw angle value is given as a set point to the controller. In this case, along with achieving the yaw angle, only the vertical motion of the rotor is expected. The simulated response for PID and LQR controller is shown in Fig.8 and Fig.9 respectively. The settling time for yaw angle for the PID controller is 0.9 sec and that of LQR is 1.8sec making PID response faster. However, PID controller shows an undesired spike in the velocity curve whereas the LQR response is a gradually decreasing curve settling around 1.8sec. This drastic behavior would create instability which is undesired. Lesser the overshoot, lesser the fluctuation and more the stability.

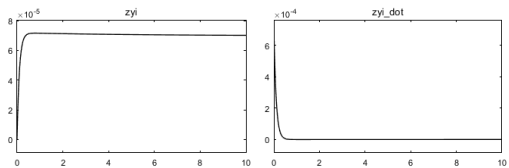

Fig. 8: PID controller response for yaw motion

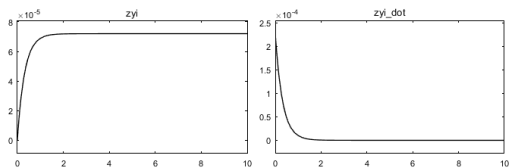

Fig. 9: LQR response for yaw motion

## VII. CONCLUSION

The two controllers show similar response with respect to settling time to achieve the desired response. Although, there are some points which give the optimal controller an advantage over the classical control method. In all the cases, 2 overshoot peaks can be seen in the PID controller output whereas only a single overshoot peak is obtained in LQR response. This means that the PID controller shows aggressive behavior as compared to the LQR controller response. Secondly, all the 6 PID controllers need to be tuned to obtain an efficient response which makes it a tedious work whereas only 2 matrices need to be tuned in case of the LQR controller. PID controller requires 6 feedback loops making the computation complex than the single control loop in case of an LQR controller. Thus, it can be concluded that LQR controller is better suited for Quadrotor control mechanism than the classical PID controller in terms of output, complexity, and, computation time.

ACKNOWLEDGMENT

We would like to express our gratitude towards the staff of the Control Systems Laboratory at BITS Pilani, Hyderabad Campus for their assistance in making this project a reality.